\documentclass[10pt]{article}
\usepackage[dvips]{graphicx}
\begin{document}

\begin{center}
\textbf{Voltage dependence of rate functions for  Na+  channel inactivation within a membrane}
 \end{center}
\begin{center} S. R. Vaccaro  \end{center}
\begin{center} 
{\em Department of Physics, University of Adelaide, Adelaide, South Australia, 
5005, 
Australia} \\
  \end{center}
{\em svaccaro@physics.adelaide.edu.au} \\

{\bf Abstract}

\begin{quotation}
 The inactivation of a Na+ channel occurs when the activation of the charged S4 segment of domain DIV is
 followed by the binding of an intracellular hydrophobic motif  which blocks conduction through the ion pore.
 The voltage dependence of Na+ channel inactivation is, in general, dependent on the rate functions of the S4
 sensors of  each of the domains DI to DIV. If the activation of a single voltage sensor that 
regulates the  Na+ channel  conductance is coupled to a two-stage inactivation process, 
the  rate functions for inactivation and recovery from inactivation, as well as 
the time dependence of the Na+ current in terms of the variables m(t) and h(t), may be derived from a
 solution to the master equation for interdependent  activation and inactivation.
 The rate functions have a voltage dependence that is consistent with the Hodgkin-Huxley empirically determined 
  expressions,  and exhibit saturation for both depolarized and hyperpolarized clamp potentials.
\end{quotation}
\newpage

{\bf INTRODUCTION}  

 The opening and subsequent inactivation of Na+ channels and the activation of K+ channels generate
 the action potential in nerve and muscle membranes \cite{hh}. The Na+ channel transient current during a
 depolarizing voltage clamp may be described by the expression $m^3 h$ where the activation variable $m$ 
and inactivation variable $h$ satisfy first order rate equations with rate functions dependent on the 
potential difference across the membrane. Support for the assumption that activation and inactivation 
are separate processes was provided by the removal of Na+  inactivation from the squid axon membrane 
by the  internal perfusion of pronase without affecting activation kinetics \cite{cbr}. However,  there 
is a delay in the onset of  Na+ channel inactivation that is dependent on the  time-course of
channel activation, and  Na+ channel inactivation partially immobilizes the gating 
charge associated with activation, and it was assumed  that the voltage dependence of 
inactivation was derived from the Na+ activation process  \cite{ab}. There is also a delay in the recovery from inactivation  that is dependent on the time-course of deactivation, and the rate of recovery from inactivation saturates for large hyperpolarizing potentials \cite{kb}, and therefore, activation and inactivation are interdependent or coupled processes.

The Na+ channel protein is comprised of four domains DI to DIV, each containing  six alpha-helical segments 
S1 to S6, and in each domain the voltage sensor, the S4 segment,  has positively charged residues located at
 every third position. The re-entrant loop between S5 an S6 forms the ion-selective filter at the extracellular
end of the pore, whereas the intracellular end of the pore is formed by the S6 segments. The inactivation gate 
is an IMF motif that is positioned on an intracellular loop between DIII and DIV, and interacts with and blocks
the flow of ions through the inner mouth of the pore \cite{wpswgc}. Based on voltage clamp fluorometry, 
in response to membrane depolarization, the transverse motion of the charged S4 segments of the 
Na+ channel domains  DI to DIII is associated with activation, whereas the slower movement 
of DIV S4 is correlated with inactivation   \cite{crgfb,cb}.  This may occur for small depolarizations 
when the ion  channel is usually closed (closed-state inactivation) or for larger depolarizations when the 
S4 segments of the domains D1 to D3 are activated (open-state inactivation).

In a naturally occurring paramyotonia congenita mutation of the outermost arginine residue of DIV S4 in the human 
muscle Na+ channel,  the inactivation rate is decreased with little voltage dependence
 for moderate depolarizations \cite{cgh,ch}, and therefore,  the voltage dependence of inactivation is dependent on 
charged residues in the S4 segment of the DIV domain. The voltage dependence of 
the open to inactivated transition was also demonstrated by comparison of gating current measurement in wild-type 
and ApA toxin modified cardiac Na+ channels \cite{sh}. By measurement of the OFF gating charge during repolarization 
in an inactivation modified mutant of the human heart Na+ channel, it was estimated that the  
DIV S4 sensor contributes approximately 30\% to the OFF charge, approximately 20\% may be attributed to 
the DIII S4 sensor when the inactivation gate is intact, and the rate-limiting step is the 
motion of the DIV S4 sensor and not the unbinding of the inactivation gate \cite{skh}.

In order to account for the effect of double-cysteine mutants of S4 gating charges on the ionic current of
 the bacterial Na+ channel NaChBac, it has been proposed that at least two transitions are required during the activation of
 each voltage sensor  \cite{dystc}. This conclusion is consistent with an earlier result that 
cross-linking a DIV S4 segment from the extracellular surface inhibits inactivation during 
membrane depolarization whereas cross-linking the same  segment from the inside inhibits 
activation of the Na+ channel, and therefore, the DIV S4 sensor translocates across the membrane 
in two stages  \cite{hdb}. A Na+ channel model that assumes that the motion of  the DIV S4 sensor
 includes a first stage that is necessary for opening of the channel, and a second stage that 
is required for inactivation,  provides a good description of gating and ionic currents \cite{arm1}.
 The measurement of gating currents for charge neutralized 
segments in each domain of the Na+ channel gives additional support to the conclusion that 
the two stage activation of the DIV S4 sensor is correlated with ion channel inactivation \cite{cgabc}.

 In this paper,  expressions for the voltage dependence of the rate of inactivation and recovery from inactivation
 are derived by assuming that  Na+ channel inactivation  is a two stage process, where 
the activation of DIV S4 is correlated with the binding of the inactivation motif to the ion pore. However, Na+ channel inactivation is, in general, dependent on the rate functions of the S4 sensors of each of
 the domains DI to DIV, and from a solution of the master equation for the activation of 
 Na+ channel  conductance by a single voltage sensor that is coupled to a two-stage inactivation process,
 it is  shown  that the voltage dependence of  the rate functions for inactivation and recovery from inactivation  have a similar form  to empirical expressions for Na+ channels  \cite{hh,kb}, and in particular, the exponential variation exhibits  saturation for both depolarized and hyperpolarized clamp potentials.

{\bf INDEPENDENT ACTIVATION AND INACTIVATION OF A Na CHANNEL}

Inactivation  of a Na+ channel may be described as the transverse motion of the charged S4 segment of the domain DIV, with rate functions $\alpha_{i}$ and $\beta_{i}$, followed by the binding of an intracellular hydrophobic motif  which blocks conduction through the ion pore, with rate functions $\gamma_{i}$ and $\delta_{i}$  
 (see Fig. 1). Assuming that the transition of the DIV S4 segment across two potential barriers occurs within an energy landscape, it may be shown from a solution of the Smoluchowski equation \cite{vac,vac1,vac2} 
that the occupation probabilities of the permissive states $h_1$, $h_2$ and the inactivated state  $h_3$ are determined by

\begin{eqnarray}
\frac{dh_{1}}{dt} & = & -\alpha_i h_{1}(t)+\beta_i h_{2}(t), \label{i1} \\
\frac{dh_{2}}{dt} & = & \alpha_i h_{1}(t)+\delta_i h_3(t)-(\beta_i +\gamma_i)h_{2}(t), \label{i2} \\
\frac{dh_{3}}{dt} & = & \gamma_i h_{2}(t)-\delta_i h_3(t),  \label{i3}
\end{eqnarray}
where each rate is an exponential function of  the membrane potential.
    If the Na+ channel is depolarized to a clamp potential $V$  from a large hyperpolarized holding potential ($h_{1}(0)=1$, and $h_{2}(0)=h_{3}(0)=0$), and it is assumed that the first forward and backward transitions are rate-limiting \cite{lcfb,lb}  
( $\beta_i \gg \delta_i $ and $\gamma_i \gg  \alpha_i$),  the solution of Eqs. (\ref{i1})  to (\ref{i3}) is 
 \cite{vac3}
\begin{equation}	
h_{3I}(t) = \frac{\alpha_i \gamma_i }{\omega_{1}\omega_{2}}+\frac{\alpha_i \gamma_i }{
\omega_{1}(\omega_{1}-\omega_{2})}\exp (-\omega_{1}t)-\frac{\alpha_i
\gamma_i }{\omega_{2}(\omega_{1}-\omega_{2})}\exp (-\omega_{2}t).  \label{inact2} 
\end{equation}
where $\omega _{1}\approx (\gamma_i \alpha_i +\delta_i (\alpha_i +\beta_i ))/(\gamma_i +\beta_i)$ 
and $\omega_{2}\approx \gamma_i +\beta_i \gg \omega_{1}$.
 
However, if the DIV S4 sensor is initially in the inactivated state ($h_{1}(0)=h_{2}(0)=0$, and $h_{3}(0)=1$), and the  membrane is hyperpolarized to a potential $V$, it may be shown that

\begin{equation}	
h_{3D}(t) =  \frac{\alpha_i \gamma_i + \delta_i (\alpha_i +\beta_i )\exp (-\omega_{1}t)}
{\alpha_i \gamma_i + \delta_i (\alpha_i +\beta_i )}.  \label{deinact2} 
\end{equation}
Eqs.  (\ref{inact2})  and  (\ref{deinact2}) are solutions of the rate equation \cite{hh}
\begin{equation}
\frac{dh_3}{dt}= \beta_{h,2}-(\alpha _{h,2}+\beta_{h,2})h_3,  \label{rateh1}
\end{equation}

\noindent where
\begin{eqnarray}
\alpha_{h,2}(V) & \approx & \frac{\delta_i (\alpha_i + \beta_i)}{\gamma_i + \beta_i},  \label{inalf1} \\
\beta_{h,2}(V)  & \approx &   \frac{\alpha_i \gamma_i}{\gamma_i + \beta_i}.  \label{inbet1}
\end{eqnarray}	

\noindent 
Therefore, the probability of the permissive states $h = h_1 + h_2$  satisfies

\begin{equation}
\frac{dh}{dt}= \alpha_{h,2}-(\alpha_{h,2}+\beta_{h,2})h,  \label{rateh2}
\end{equation}
and Eqs. (\ref{inalf1}) and (\ref{inbet1}) provide a good fit to the empirical inactivation rate functions  
$\alpha_{h}$ and $\beta_{h}$  for the squid axon Na+ channel  \cite{hh} (see Fig. 2).

    Based on the measurement of a rising phase of the gating current in a squid axon membrane and the chemical structure of a Na channel, at least two transitions are required for the activation of each voltage sensor \cite{dystc,ke}. Therefore, for each voltage sensor from domains DI to DIII, assuming no cooperativity between sensors,  the occupation probabilities of the  closed  states $m_1$, $m_2$ and the open state $m$ (see Fig. 4)
 are determined by 
\begin{eqnarray}
\frac{dm_{1}}{dt} & = & -\alpha_a m_{1}(t)+\beta_a m_{2}(t), \label{mm1} \\
\frac{dm_{2}}{dt} & = & \alpha_a m_{1}(t)+\delta_a m(t)-(\beta_a +\gamma_a)m_{2}(t), \label{mm2} \\
\frac{dm}{dt} & = & \gamma_a m_{2}(t)-\delta_a m(t).  \label{mm}
\end{eqnarray}

\noindent 
where the transition rates $\alpha_{a}$,  $\beta_{a}$, $\gamma_a$ and $\delta_a$ are exponential
 functions of  the membrane voltage  $V$.    If we assume that $\beta_a \gg \delta_a $ and 
$\gamma_a \gg  \alpha_a$ \cite{lcfb,lb}, from the solution of Eqs. (\ref{mm1}) to  (\ref{mm}) during activation ($m_1(0)=1$), and deactivation ($m(0)=1$), 
 it may be shown that the open state $m$ may be approximated by

\begin{eqnarray}
m_{A}(t) & \approx & \frac{\alpha_a \gamma_a }{\alpha_a \gamma_a +\delta_a (\alpha_a +\beta_a )}
[ 1-\exp (-\omega_{1}t)] ,  \label{act2} \\
m_{D}(t)  & \approx & \frac{\alpha_a \gamma_a +\delta_a (\alpha_a +\beta_a )\exp (-\omega
_{1}t)}{\alpha \gamma_a +\delta_a (\alpha_a +\beta_a )}.  \label{deact2}
\end{eqnarray}
where the low frequency $\omega_{1} \approx (\gamma_a \alpha_a + \delta_a (\alpha_a +\beta_a ))/(\gamma_a +\beta_a )$. 
Eqs. (\ref{act2}) and (\ref{deact2})   satisfy the rate equation \cite{hh}
\begin{equation}
\frac{dm}{dt}= \alpha_{m,2}-(\alpha_{m,2}+\beta_{m,2})m,  \label{ratem2}
\end{equation}

\noindent and the rate functions

\begin{eqnarray}
\alpha_{m,2}(V) & \approx &  \frac{\alpha_a }{1+\beta_a/\gamma_a},  \label{alf2} \\
\beta_{m,2}(V)  & \approx &  \frac{\delta_a (\alpha_a +\beta_a)}{\gamma_a +\beta_a},  \label{bet2}
\end{eqnarray}

\noindent provide a good fit to the empirical functions   $\alpha_{m}$ and $\beta_{m}$   for the squid axon Na channel \cite{hh} (see Fig. 4). However, the cooperativity between the S4 sensors in domains DI to DIII also contributes to the voltage dependence of the Na+ conductance rate functions, and more recent models have adopted 
exponential functions for both $\alpha_{m}$ and $\beta_{m}$ \cite{kb}. 
%\newpage

{\bf COUPLED MODELS OF ACTIVATION AND INACTIVATION OF A Na CHANNEL
}       

  The time-dependence of the Na+ current in the squid axon may be expressed as $m^3 h$
where the activation variable m(t) and inactivation variable h(t)  satisfy the rate equations \cite{hh}
\begin{equation}
\frac{dm}{dt}= \alpha_{m}-(\alpha_{m}+\beta_{m})m,  \label{ratem2}
\end{equation}
\begin{equation}
\frac{dh}{dt}= \alpha_{h}-(\alpha_{h}+\beta_{h})h.  \label{rateh2}
\end{equation}
The Hodgkin-Huxley (HH) description of the Na current  is equivalent to an 8-state master equation where three independent  voltage sensors may activate, and inactivation may occur from the
open state or from each  of the closed states \cite{arm}.  In this section, we assume that the activation of a single voltage sensor  regulating the  Na channel  conductance is coupled to a two-stage inactivation process
 (see Fig. 5), and therefore, the kinetics may be described by a master equation where the occupation
 probabilities of the closed states $C_1$ and $A_1$,  the open states $O$ and $A_2$, and  the inactivated (or blocked) states $B_1$ and $B_2$ are determined by
\begin{eqnarray}
\frac{dC_1}{dt} & = &  -(\alpha_1 + \alpha_O)C_1(t)  + \beta_O O(t) + \beta_1 A_1(t) \label{6c1} \\
\frac{dO}{dt} & = &  \alpha_O C_1(t) - (\beta_O + \alpha_2)O(t) + \beta_2 A_2(t) \label{6o} \\
\frac{dA_1}{dt} & = &
 \alpha_1 C_1(t) -(\alpha_A + \beta_1 + \gamma_1)A_1(t) + \delta_1 B_1(t) + \beta_A A_2(t) \label{6a1} \\
\frac{dA_2}{dt} & = & \alpha_2 O(t) -(\beta_A + \beta_2 + \gamma_2)A_2(t) +\delta_2 B_2(t) + \alpha_A A_1(t) \label{6a2} \\
\frac{dB_1}{dt} & = & \gamma_1 A_1(t)-(\alpha_B+ \delta_1) B_1(t) +\beta_B B_2(t) \label{6b1} \\
\frac{dB_2}{dt} & = &\gamma_2 A_2(t) +\alpha_B B_1(t) -(\beta_B+ \delta_2) B_2(t), \label{6b2}
\end{eqnarray}
and the transition rates are functions of the membrane voltage $V$. 

Assuming that the first forward and  backward transitions for inactivation are rate limiting, 
 $\beta_j \gg \delta_j$ and  $\gamma_j \gg \alpha_j$, for $j=1,2$,  $A_1$ and $A_2$ satisfy
\begin{eqnarray}
A_1  & \approx &  \frac{\alpha_1 C_1 + \delta_1 B_1}{\beta_1 + \gamma_1},  \\
A_2  & \approx & \frac{\alpha_2 O + \delta_2 B_2}{\beta_2 + \gamma_2},  
\end{eqnarray}

\noindent and therefore, Eqs. (\ref{6c1}) to (\ref{6b2})  may be reduced to a four state master equation 
(see Fig. 6)
\begin{eqnarray}
\frac{dC_1}{dt} & = &  -(\rho_1 + \alpha_O)C_1(t) + \beta_O O(t)+ \sigma_1 B_1(t) \label{4c1} \\
\frac{dO}{dt}   & = &   \alpha_O C_1(t) - (\beta_O + \rho_2)O(t) +\sigma_2 B_2(t)  \label{4o} \\
\frac{dB_1}{dt} & = &  \rho_1 C_1(t) -(\alpha_B+ \sigma_1) B_1(t)+\beta_B B_2(t)  \label{4b1} \\
\frac{dB_2}{dt} & = &  \rho_2 O(t) + \alpha_B  B_1(t) -(\beta_B+ \sigma_2) B_2(t), \label{4b2} 
\end{eqnarray}
where the derived inactivation rate functions
\begin{eqnarray}
\rho_1    & \approx &  \frac{\alpha_1 \gamma_1}{\beta_1 + \gamma_1}, 
\rho_2    \approx   \frac{\alpha_2 \gamma_2}{\beta_2 + \gamma_2},  \\ \label{rhosg1}
\sigma_1  & \approx & \frac{\delta_1 (\beta_1 + \alpha_1)}{\beta_1 + \gamma_1}, 
\sigma_2   \approx  \frac{\delta_2 (\beta_2 + \alpha_2)}{\beta_2 + \gamma_2}. \label{rhosg2}
\end{eqnarray}

If  Na+ activation and inactivation are independent processes ($\alpha_O = \alpha_B$, $\beta_O = \beta_B$,  
$\rho_1 = \rho_2 = \rho$   and $\sigma_1 =\sigma_2 = \sigma$), and the Na+ channel is depolarized to a clamp potential  $V$  from a large hyperpolarized holding potential, the open state 
$O(t) = m(t)h(t)$  where the activation and inactivation variables
\begin{eqnarray}
m(t) & = & \frac{\alpha_O}{\alpha_O + \beta_O} \left( 1-\exp [-(\alpha_O + \beta_O)t] \right) ,  \label{msol} \\
h(t)  & = & \frac{ \sigma + \rho \exp (-(\rho + \sigma)t)}{\sigma + \rho},  \label{hsol}
\end{eqnarray}
and satisfy the rate equations (\ref{ratem2})  and (\ref{rateh2})  when $\alpha_{O} = \alpha_{m}$,
 $\beta_O = \beta_m$, $\rho  = \beta_h$  and $ \sigma  = \alpha_h$ (see Fig. 7).
 Similarly, if the Na+ channel is
 hyperpolarized to a clamp potential  $V$  from a large
 depolarized holding potential, $C_1(t) = (1 - m(t))h(t)$ (see Fig. 8) where 
\begin{eqnarray}
m(t) & = & \frac{\alpha_O + \beta_O \exp (-(\alpha_O + \beta_O)t)}{\alpha_O + \beta_O},  \label{msol2} \\
h(t)  & = & \frac{ \sigma}{\rho + \sigma} (1 - \exp [-(\rho + \sigma)t]).  \label{hsol2}
\end{eqnarray}
In the HH model, the inactivation rate functions are assumed to be independent of the Na+ conductance activation
functions but, by contrast, the  rate of recovery for inactivation in hippocampal neurons is also dependent  on the Na+ 
channel deactivation functions \cite{kb}. 

More generally, the Na+ current during deactivation of the channel is very small \cite{ab}, and therefore, 
the deinactivation rate  $\sigma_2 \approx 0$. For a  depolarizing clamp potential $V$  from a large 
hyperpolarized holding potential,   the solution of Eqs. (\ref{4c1})  to   (\ref{4b2}) when 
$\rho_1 = \rho_2 = \rho$ is  
\begin{eqnarray}
C_1(t) & = &  C_{1s} + \Sigma_{j = 1}^{3} k_{j+1} (\omega_{j} - \beta_O - \rho_2) \exp (-\omega_{j}t)  \\
O(t)   & = &   O_{s}  - \Sigma_{j = 1}^{3} k_{j+1} \alpha_O \exp (-\omega_{j}t)   \label{oodep} \\
B_1(t) & = & B_{1s}   + \Sigma_{j = 1}^{3} k_{j+1} X_1(\omega_{j}) \exp (-\omega_{j}t)  \\
B_2(t) & = & B_{2s}   + \Sigma_{j = 1}^{3} k_{j+1} X_2(\omega_{j}) \exp (-\omega_{j}t), 
\end{eqnarray}
where  $C_{1s} = k_1 \sigma_1 \beta_B (\beta_O + \rho)$, $O_{s} = k_1 \sigma_1 \beta_B \alpha_O$, 
 $B_{1s} = k_1 \rho \beta_B (\alpha_O + \beta_O + \rho)$, 
$B_{2s} =  k_1 \rho \alpha_B (\alpha_O + \beta_O + \rho) + \rho \alpha_O \sigma_1$, 
$k_1^{-1} = [(\alpha_O + \beta_O + \rho)(\sigma_1 \beta_B +  \rho(\alpha_B+\beta_B))+ \rho \sigma_1 \alpha_O]$,
\begin{eqnarray}
k_2 & = &  \frac{1 - k_1 \sigma_1 \beta_B \omega_{2} +k_4 (\omega_{2} - \omega_{3})}{\omega_{1}-\omega_{2}} \\
k_3 & = &  \frac{1 - k_1 \sigma_1 \beta_B \omega_{1} +k_4 (\omega_{1} - \omega_{3}))}{\omega_{2}-\omega_{1}} \\
k_4 & = & \frac{-B_{1s} (\omega_{2}-\omega_{1}) + r_2 X_1(\omega _{1}) - r_1 X_1(\omega _{2})}
{(\omega_{2}-\omega_{1}) X_1(\omega _{3})+ (\omega_{3}-\omega_{2}) X_1(\omega _{1}) - 
(\omega_{3}-\omega_{1}) X_1(\omega _{2})  }, \\
X_1(\omega) & = &  \frac{ -\rho \alpha_O \beta_B  - \rho (\omega - \beta_B)(\omega - \beta_O -  \rho)}
{\omega^2 - \omega (\alpha_B  + \beta_B + \sigma_1) + \sigma_1 \beta_B} ,  \\
X_2(\omega) & = &  \frac{  -\rho \alpha_B(\alpha_O + \beta_O + \rho)  - \rho \alpha_O \sigma_1 +
 \rho \omega (\alpha_O  + \alpha_B )}
{\omega^2 - \omega (\alpha_B  + \beta_B + \sigma_1) + \sigma_1 \beta_B}.
\end{eqnarray}
For depolarization clamp potentials,    $\omega_2 \approx \alpha_O + \beta_O + \rho$, 
$\omega_3 \approx \alpha_B + \beta_B + \sigma_1$,
and
\begin{equation}
 \omega_1 \approx \alpha_h + \beta_h,
\end{equation}
\noindent  where the inactivation rate 
\begin{equation}	
\beta_h =  \frac{\rho}{1 + \sigma_1/(\alpha_B + \beta_B)} +  
\frac{\rho \sigma_1 \alpha_O}{(\alpha_O + \beta_O + \rho)(\alpha_B + \beta_B + \sigma_1)},  \label{betah}
\end{equation}
and the rate of recovery from inactivation
\begin{equation}	
\alpha_h =  \frac{\beta_B  \sigma_1}{\alpha_B + \beta_B + \sigma_1}.  \label{alphah}
\end{equation}
The second term in Eq.(\ref{betah}) only makes a contribution to $\beta_h$  for small clamp potentials.
\noindent   If $\alpha_1 = \alpha_2$ and $\delta_1$ are independent of $V$, $\omega_1$ saturates for both 
large positive and negative clamp potentials (see Fig. 9). From Eq. (\ref{oodep}), as $k_4 \approx 0$
 for a depolarizing potential, we may write
\begin{eqnarray}	
O(t) & \approx & \frac{\alpha_O}{\alpha_O + \beta_O}
 \left( \frac{\alpha_h (1 - \exp [-(\alpha_O + \beta_O+\rho)t] )}{\alpha_h + \beta_h} + \right. \\
     &  & \left. \frac{\beta_h  \exp [-(\alpha_h + \beta_h)t](1 - \exp [-(\alpha_O + \beta_O)t] )}{\alpha_h + \beta_h} \right),
\end{eqnarray}
and therefore, $O(t) \approx m(t) h(t)$  (see Fig. 10) where $m(t)$ is defined in Eq.  (\ref{msol}) and 
\begin{equation}
h(t) =  \frac{\alpha_h + \beta_h \exp (-(\alpha_h + \beta_h) t)}{\alpha_h + \beta_h}.  \label{hsol2}
\end{equation}
That is, the HH description of  the Na+ current in terms of the variables $m(t)$ and $h(t)$ 
 is an approximation to an expression that may be derived from a solution to a coupled model
of Na+ activation and two stage inactivation for which the the deinactivation rate  $\sigma_2 \approx 0$.

For a moderate hyperpolarizing clamp potential from a depolarized holding potential, the inactivation rates $\rho_1, \rho_2 \approx 0$, and the solution of 
Eqs. (\ref{4c1}) to (\ref{4b2}) is (see Fig. 11)
\begin{eqnarray}
C_1(t) & = &  \frac{\beta_O }{\alpha_O + \beta_O}  + 
\frac{Y_1(\omega _{1}) }{\omega _{1}-\omega _{2}}\exp (-\omega _{1}t) 
  - \frac{Y_1(\omega_{2}) }{\omega _{1}-\omega _{2}}\exp (-\omega_{2}t)+  \nonumber \\
    &   &  \frac{\alpha_O \omega_{1} \omega_{2} + \omega_{3}\sigma_2 (\beta_O - \alpha_B -\sigma_1) } 
{\omega_{3}(\omega_{1}-\omega_{3})(\omega_{2}-\omega_{3})}\exp (-\omega_{3}t) \label{c1h} \\
O(t)   & = & \frac{\alpha_O}{\alpha_O + \beta_O}  +
\frac{Y_2(\omega _{1}) }{\omega _{1}-\omega _{2}}\exp (-\omega _{1}t) - 
\frac{Y_2(\omega_{2}) }{\omega _{1}-\omega _{2}}\exp (-\omega_{2}t)-  \nonumber \\
    &   &  \frac{\alpha_O \omega_{1}\omega_{2} + \omega_{3}\sigma_2 (\beta_O - \alpha_B -\sigma_1)}
{\omega_{3}(\omega_{1}-\omega_{3})(\omega_{2}-\omega _{3})}\exp (-\omega_{3}t) , \label{ooh} \\
B_1(t) & = & -\frac{\beta_B }{\omega _{1}-\omega _{2}}\exp (-\omega _{1}t) + 
\frac{\beta_B}{\omega _{1}-\omega _{2}} \exp (-\omega_{2}t), \label{b1h} \\
B_2(t) & = & \frac{\omega_{1}- \alpha_B -\sigma_1}{\omega_{1}-\omega_{2}}\exp (-\omega_{1}t)-  
    \frac{\omega_{2}- \alpha_B -\sigma_1}{\omega _{1}-\omega_{2}} \exp (-\omega_{2}t),  \label{b2h} 
\end{eqnarray}
where $\omega_3 = \alpha_O + \beta_O$, and  $\omega_1, \omega_2$ are solutions of the characteristic equation
\begin{equation}	
\omega^2 - \omega (\alpha_B + \beta_B + \sigma_1 + \sigma_2) +  \sigma_1 \beta_B  +
  \sigma_2(\alpha_B + \sigma_1) = 0, \label{char2}
\end{equation}
\begin{eqnarray}
Y_1(\omega) & = & \frac{  - \beta_O \omega_{1}\omega_{2}   + \omega (\sigma_1 \beta_B + \sigma_2 \beta_O)}
{\omega(\omega- \alpha_O  - \beta_O)},   \\
Y_2(\omega) & = &  \frac{ \omega_{1}\omega_{2}(\sigma_2 - \alpha_O ) - \omega \sigma_2 (\beta_B + \sigma_2 - \alpha_O)}
{\omega(\omega - \alpha_O  - \beta_O)}. 
\end{eqnarray}

If $\rho_1 = \rho_2$, from the application of microscopic reversibility to
the four state system (see Fig. 6), $\beta_B =  \beta_O \alpha_B \sigma_2/\alpha_O \sigma_1 \ll \beta_O$ for 
$\sigma_1 \gg \sigma_2 \approx 0$. From Eq. (\ref{char2}), assuming that  $\alpha_B \ll \beta_B \ll \sigma_1$
for a small  hyperpolarizing potential, the lowest frequency  $\omega_1 \approx \beta_B$,
 whereas for $ \beta_B \gg \sigma_1$, $\omega_1 \approx \sigma_1$
 (see Fig. 9). The conclusion that, for small hyperpolarizing potentials,  the recovery rate for inactivation 
$\alpha_h \approx \beta_B \propto \beta_m$ is supported by the HH data where
$\alpha_h(V)$  and $\beta_m(V)$ have a similar voltage dependence and 
$\beta_m(V) \approx 57 \alpha_h(V)$  \cite{hh}.  However, if $\alpha_B$
is an exponential function of $V$  such that $\alpha_B + \beta_B \gg \sigma_1$ \cite{kb} (see Fig. 12), 
from Eq. (\ref{char2}), $\omega_2 \approx \alpha_B + \beta_B + \sigma_1$, and 
\begin{equation}	
 \omega_1 \approx  \frac{\sigma_1\beta_B}{\alpha_B + \beta_B + \sigma_1} .
\end{equation}

\noindent Therefore, the voltage dependence  of the  rate of recovery from inactivation 
is  determined by the deinactivation rate  $\sigma_1$ and the  Na+ conductance deactivation 
functions \cite{kb}.  For small  hyperpolarizing potentials ($\beta_B \ll \alpha_B$), 
$\omega_1 \approx \sigma_1 \beta_B/(\alpha_B + \sigma_1) $, and  may be approximated by an exponential 
function of $V$  \cite{hh} but saturates at a more negative potential when 
$\beta_B \gg \sigma_1 \gg \alpha_B$ (see Fig. 12).

From Eq. (\ref{c1h}), we may write 
\begin{equation}	
 C_1(t)  \approx   \frac{\beta_O }{\alpha_O + \beta_O} 
\left(1 - \exp (-\omega_{1}t)\left[1 +  
\frac{\omega_{1}(1 - \exp [-(\omega_{2} - \omega_{1})t])}{\omega_{2} - \omega_{1}}\right] \right) \label{c1} . 
\end{equation}

\noindent and therefore, $dC_1/dt(0) = 0$ and there is a delay in the recovery from inactivation \cite{kb,cgabc}. However, 
for large negative potentials,  $\omega_{2} \approx \beta_B \gg \omega_{1} \approx \sigma_1$, and 
Eq. (\ref{c1}) reduces to the HH expression
\begin{equation}	
 C_1(t)  \approx  [1 - m_s] h(t),  
\end{equation}

\noindent where $m_s  = \alpha_O/(\alpha_O + \beta_O)$ and  $h(t)   =  1 - \exp (-\omega_{1} t)$.

{\bf CONCLUSION}

 Hodgkin and Huxley described the voltage dependence of the Na+ channel inactivation rate and the 
rate of recovery by exponential functions which for the inactivation rate saturates for a moderate 
depolarizing potential  \cite{hh}. Based on the absence of a gating current that corresponds to
 the time course of the inactivated Na+ current, it was assumed that the transition rates governing inactivation were voltage-independent, and that the macroscopic voltage dependence of inactivation derived from 
the Na+ channel activation process  \cite{ab}. However, the voltage dependence of the 
inactivation rate is dependent on the charge on the S4 segment residues  of the DIV domain \cite{cgh,ch,sh},
 and generally, only has a minor contribution from the Na+ conductance activation rate functions. The voltage dependence of the rate of recovery from inactivation saturates for a large hyperpolarizing potential,
 and has  been attributed to the Na+ channel deactivation rate functions \cite{kb}.

 In this paper, assuming  that  Na+ channel inactivation is a two stage process, where the activation of 
DIV S4 is correlated with the binding of the inactivation motif to the ion pore, we show that 
during a voltage clamp of the Na+ channel, the solution of the master equation for the inactivation process
 may be approximated by the solution of a rate equation. The backward transition rate is, in general,  an exponential function of the membrane potential $V$,  and the forward rate may be expressed as an exponential function for small depolarizations but approaches a saturated value for a larger clamp potential, 
reflecting the voltage independence of the rate limiting step.

If  the processes of Na activation and inactivation are independent, and the activation of a single 
voltage sensor that regulates the  Na channel  conductance is coupled to a two-stage inactivation process,
the open state probability  $O(t)$ during a  depolarizing clamp potential, derived from an analytical
solution of a four state master equation,  may be expressed as  $m(t)h(t)$ 
 where $m(t)$  and $h(t)$ satisfy rate equations for activation and inactivation, and the voltage dependence of 
the rate of inactivation  provides a good approximation to the HH function  $\beta_h$  \cite{hh}.  
However, the voltage dependence of the rate of recovery from
inactivation is dependent on the rate functions for the DIV sensor and not the Na+ channel conductance
deactivation functions, as observed experimentally in hippocampal neurons \cite{kb}.

Based on the measurement of a very small ion channel current during
 deactivation \cite{kb}, the Na+ channel deactivates before recovery from inactivation, and hence the 
deinactivation rate  $\sigma_2$ to the open state is smaller than the corresponding rate $\sigma_1$ from a 
deactivated state . A more general expression for the voltage dependence of the  rate of recovery
 from inactivation  may be determined that is approximated by an exponential function for physiological 
potentials \cite{hh}, but for more negative hyperpolarizing potentials, it approaches a 
limiting value equal to the  deinactivation rate $\sigma_1$.    The inactivation rate is, in general,
 also dependent  on the Na+ conductance activation rate functions, as well as the rate functions for
 the DIV S4 sensor, but for a moderate depolarizing potential it reduces  to a two-stage expression  
when the DIV S4 activation rate $\rho_1 = \rho_2$. 

We conclude that  general expressions for the voltage
 dependence of rate functions for inactivation  and recovery from inactivation  may be determined from 
a master equation for interdependent Na+ channel activation and inactivation, and  are consistent 
with the empirical data from the squid axon \cite{hh}, hippocampal neurons \cite{kb} and Nav1.4
ion channels \cite{cgabc}. When Na+ conductance activation is regulated by a single voltage sensor, 
the HH description of the Na+ current in terms of the activation  and inactivation variables  $m(t)$ and   $h(t)$, 
is an approximation to an expression derived from the solution to a coupled model
of Na+ activation and  two-stage  inactivation where the deinactivation rate to the open state $\sigma_2 \approx 0$.

%\newpage 

\newpage 

%Figure 1
\begin{figure*}
\begin{center}
\includegraphics[width=0.7\textwidth]{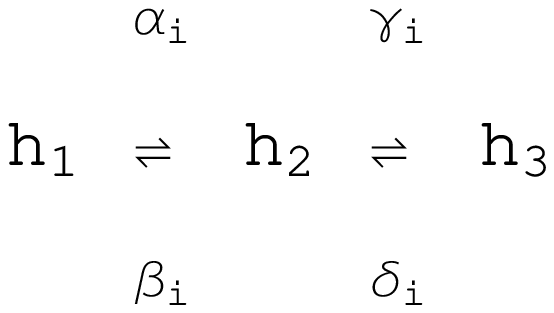}
\caption{
Inactivation model of a Na+ ion channel, where the occupation probabilities of 
the permissive states $h_1$(t), $h_2$(t), and the inactivated state $h_3$(t) 
satisfy a master equation, and the rate functions $\alpha_{i}$, $\beta_{i}$, 
 $\gamma_{i}$ and $\delta_{i}$, are voltage-dependent rate functions between states.
}
\end{center}
\end{figure*}

%Figure 2 
\begin{figure*}
\begin{center}
\includegraphics[width=0.7\textwidth]{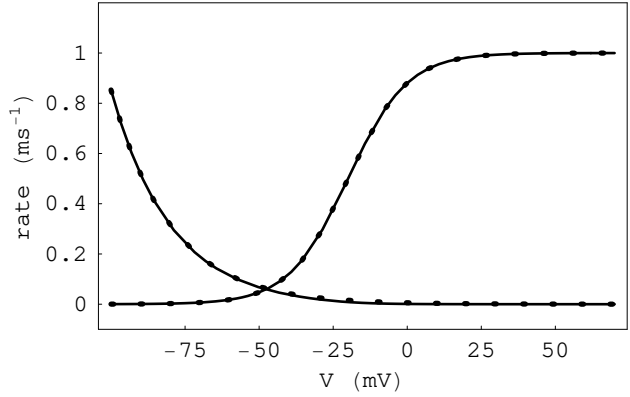}
\caption{
    The derived rate functions $\alpha_{h,2}$(V) and $\beta_{h,2}$(V) (solid line) in Eqs.  
(\ref{inalf1}) and (\ref{inbet1})  provide a good approximation to the HH
rate functions (ms$^{-1}$) $\alpha_{h} = 0.07 \exp(-(V+50)/20)$  and 
$\beta_{h} = 1/(1+ \exp(-(20+V)/10))$ (dotted line) when the inactivation rate functions 
are  $\alpha_{i}(V) = 1  \ll \gamma_{i}(V)= \exp(3)$, and 
$\beta_{i}(V)=  \exp(-2.5(V - 10)/25) \gg \delta _{i}(V)=0.07
 \exp(-(V+50)/20)$.
}
\end{center}
\end{figure*}

%Figure 3
\begin{figure*}
\begin{center}
\includegraphics[width=0.7\textwidth]{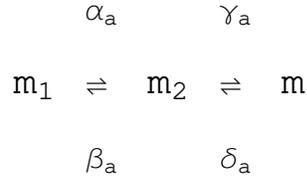}
\caption{
Activation model of a Na+ channel, where the occupation probabilities of 
the closed states $m_1$, $m_2$, and the open state $m$ satisfy a master equation, and
 $\alpha_{a}$,  $\beta_{a}$, $\gamma_{a}$ and $\delta_{a}$ are voltage-dependent rate functions
 between states.
}
\end{center}
\end{figure*}

%Figure 4 
\begin{figure*}
\begin{center}
\includegraphics[width=0.7\textwidth]{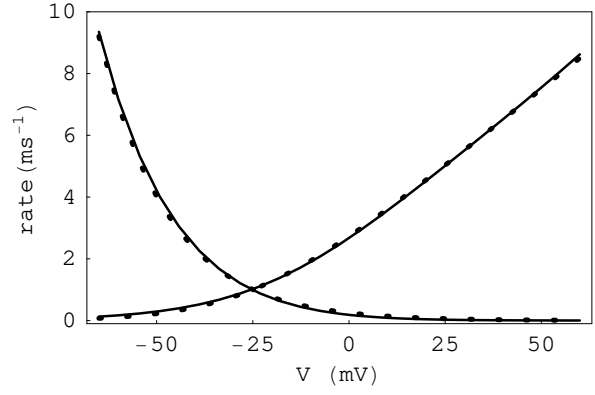}
\caption{
The derived rate functions $\alpha_{m,2}(V)$ and  $\beta_{m,2}(V)$ (solid line) 
in Eqs. (\ref{alf2}) and (\ref{bet2}) provide a good approximation to the HH rate
 functions (ms$^{-1}$) $\alpha_{m}=0.1(V+25)/(1+ \exp[-0.1(V+25)])$  and 
$\beta_{m}=4 \exp[-(V+50)/18]$ (dotted line) when the activation rate functions 
are $\alpha_{a}(V)= 2.7 \exp(0.27(V+50)/25) \ll \gamma_{a}(V)= 30 \exp(0.25(V+50)/25)$,
and   $\beta_{a}(V)= 251 \exp(-0.95(V+50)/25) \gg \delta_{a}(V) = 4.65 \exp(-0.9(V+50)/18)$.
}
\end{center}
\end{figure*}

%Figure 5 
\begin{figure*}
\begin{center}
\includegraphics[width=0.5 \textwidth]{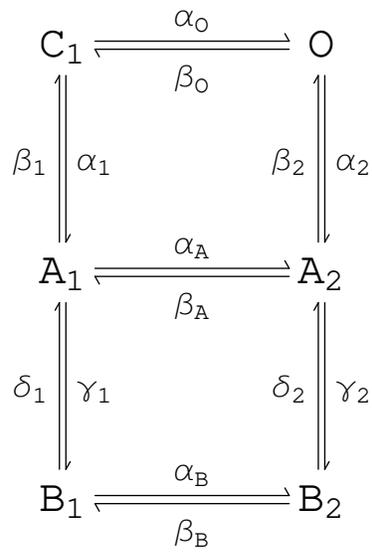}
\caption{
Six state system that describes Na+ conductance activation  between states $C_1$ and $O$, $A_1$ and $A_2$, and $B_1$ and $B_2$, is coupled to a two-stage Na+ inactivation process between states $C_{1}$ and $B_{1}$, and $O$ and $B_{2}$.
}
\end{center}
\end{figure*}

%Figure 6 
\begin{figure*}
\begin{center}
\includegraphics[width=0.5 \textwidth]{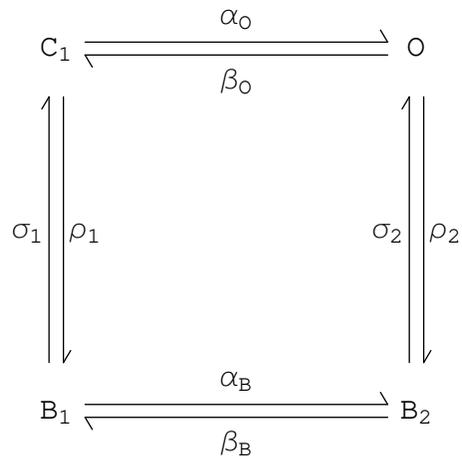}
\caption{
The six state system of Fig. 5 may be approximated by a four state system when
$\beta_j \gg \delta_j$ and  $\gamma_j \gg \alpha_j$, for $j=1,2$ where $\rho_{j}$ and $\sigma_{j}$ are 
derived rate functions for a two-stage Na+ inactivation process, defined in Eqs. (\ref{rhosg1}) and  (\ref{rhosg2}).
}
\end{center}
\end{figure*}

%Figure 7 
\begin{figure*}
\begin{center}
\includegraphics[width=0.7\textwidth]{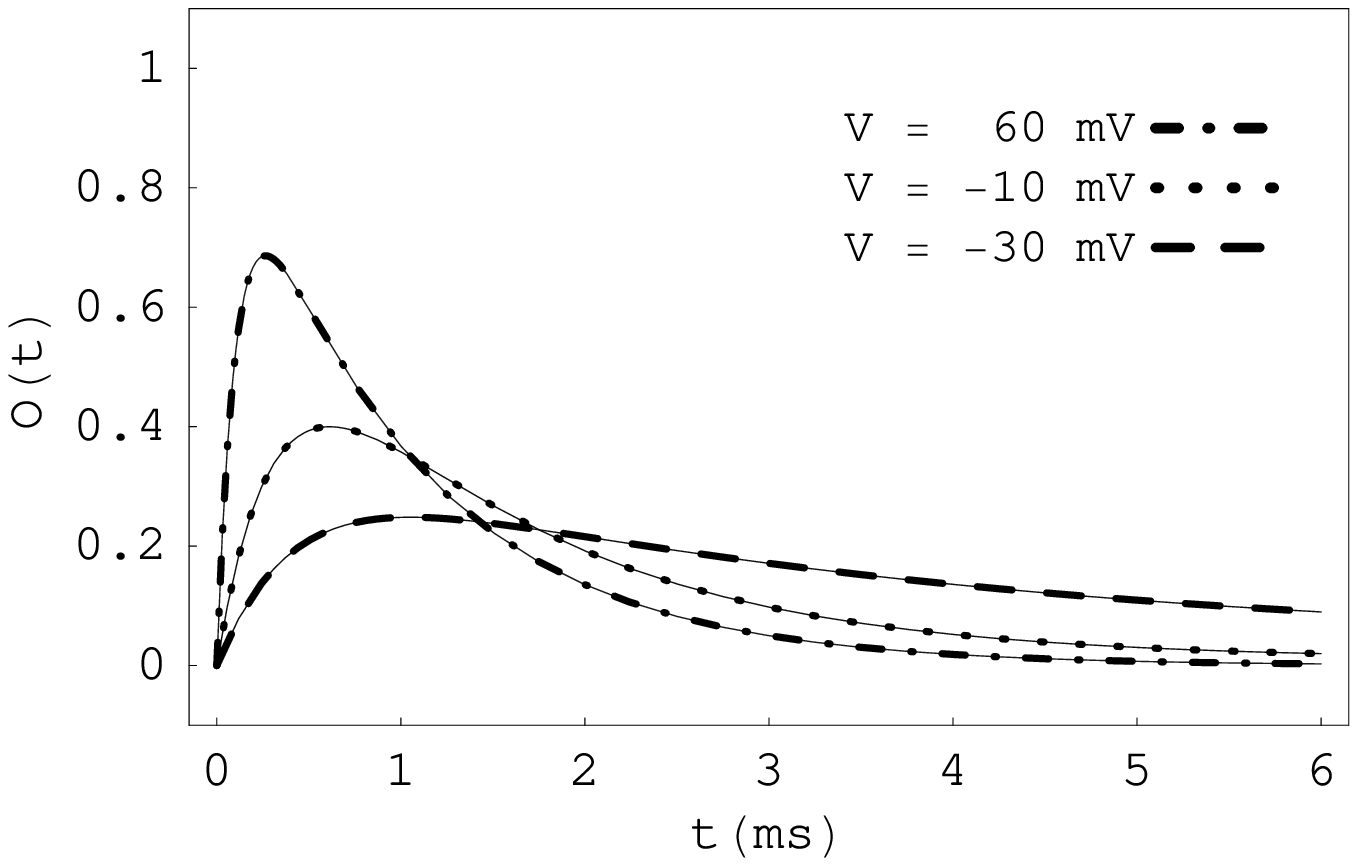}
\caption{
For a four state system where activation and inactivation are independent, the open state probability  $O(t)$
 (solid line) during a  depolarizing clamp potential, is equal to  $m(t)h(t)$ 
(dashed, dotted or dot-dashed line) where $m(t)$  and $h(t)$ are solutions of rate equations 
for activation and inactivation, and 
$\alpha_{O} = \alpha_{m} = 0.1(V+25)/(1 - \exp[-(V+25)/10])$,  $\beta_O = \beta_m = 4 \exp[-(V+50)/18]$,
 $\rho  = \beta_h = 1/(1+ \exp[-(20+V)/10])$  and   $\sigma  = \alpha_h = 0.07 \exp[-(V+50)/20]$ (ms$^{-1}$) \cite{hh}.
}
\end{center}
\end{figure*}

%Figure 8 
\begin{figure*}
\begin{center}
\includegraphics[width=0.7\textwidth]{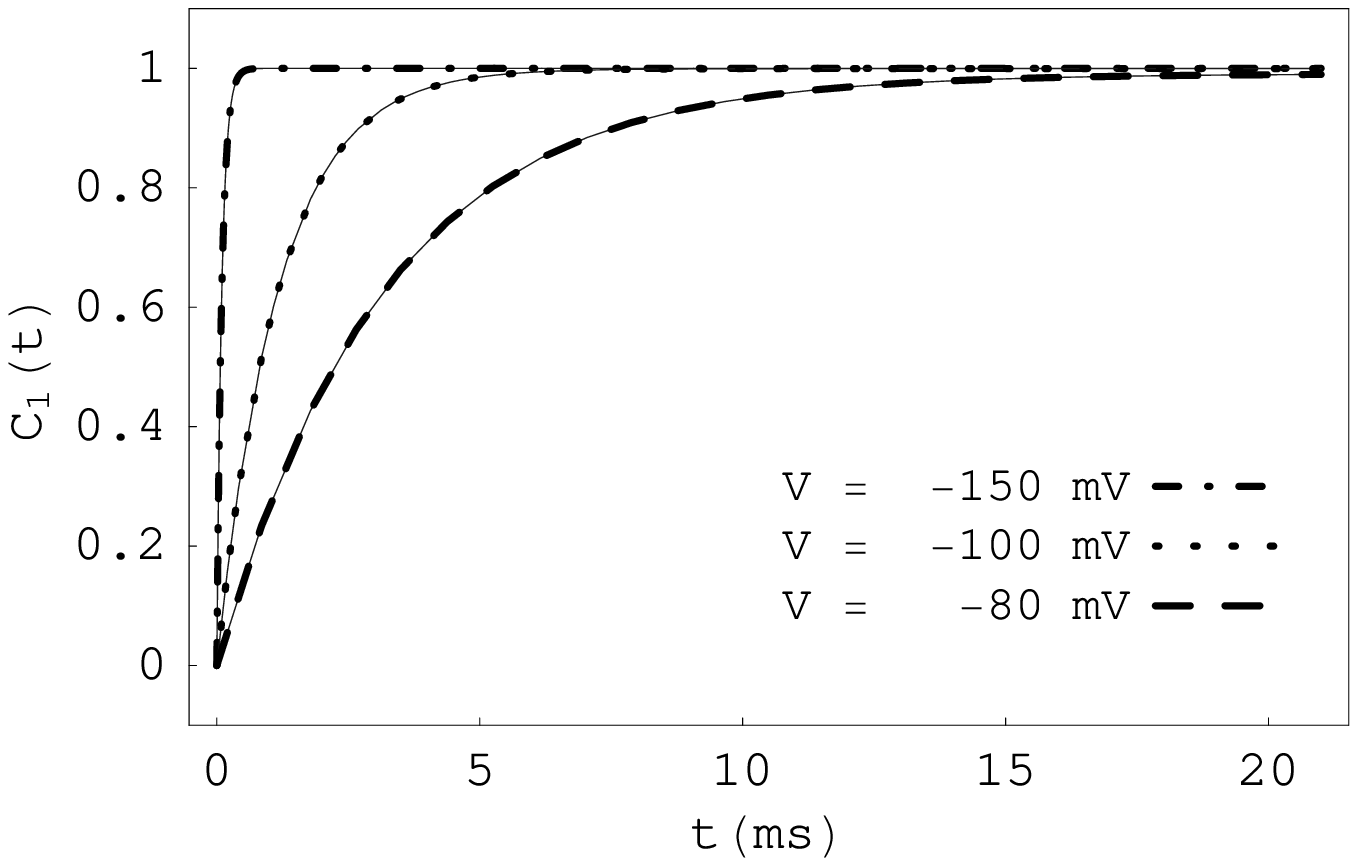}
\caption{
For a four state system where activation and inactivation are independent, the closed state probability  $C_1(t)$
 (solid line) during a  hyperpolarizing clamp potential, is equal to  $[1 - m(t)]h(t)$ 
(dashed, dotted or dot-dashed line) where $m(t)$  and $h(t)$ are solutions of rate equations 
for activation and inactivation, and 
$\alpha_{O} = \alpha_{m} = 0.1(V+25)/(1 - \exp[-(V+25)/10])$,  $\beta_O = \beta_m = 4 \exp[-(V+50)/18]$,
 $\rho  = \beta_h = 1/(1+ \exp(-(20+V)/10))$  and   $\sigma  = \alpha_h = 0.07 \exp(-(V+50)/20)$ (ms$^{-1}$).
}
\end{center}
\end{figure*}

%Figure 9 
\begin{figure*}
\begin{center}
\includegraphics[width=0.7\textwidth]{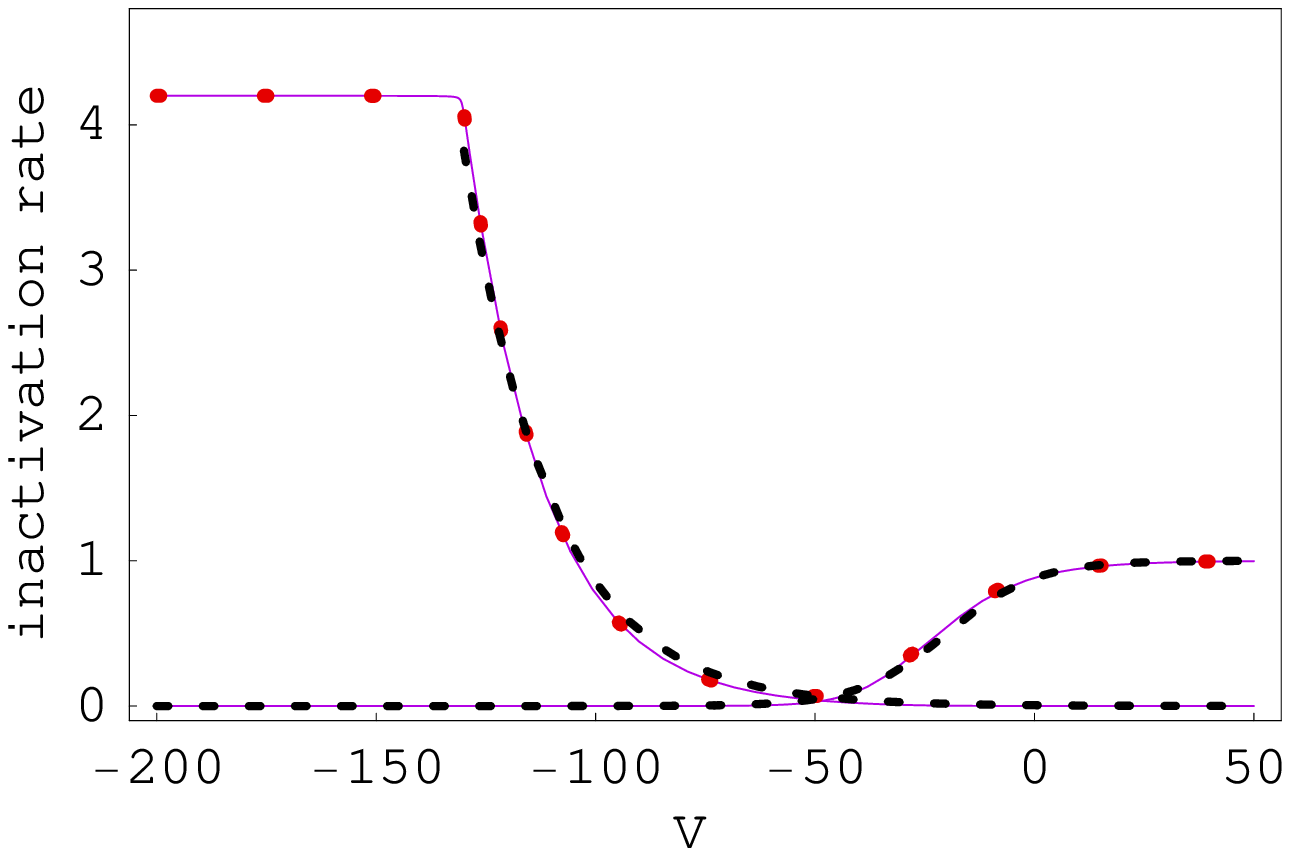}
\caption{
Voltage dependence of the HH Na+ channel inactivation rate functions $\beta_h = 1/(1+ \exp(-(20+V)/10))$ 
and $\alpha_h = 0.07 \exp(-(V+50)/20)$ (dashed line) may be approximated by analytical
 expressions in  Eqs. (\ref{betah})  and (\ref{alphah}) (solid line) derived from a master equation  for a four state system 
where activation and  two stage inactivation  are interdependent, and by the voltage dependence
 of the lowest frequency of the system determined  numerically (dotted line) where the rate 
 functions are $\alpha_{1}(V) = \alpha_{2}(V) = 1$,  
$\gamma_{1}(V)= \gamma_{2}(V) = \exp(3)$, $\beta_{1}(V)= \beta_{2}(V) =  \exp(-2(V-10)/25)$ 
$\delta_{1}(V)= 4.2$, $\delta_{2}(V)= 0$, $\alpha_{O}  = 0.1(V+25)/(1 - \exp[-(V+25)/10]) = \alpha_{B}/3$ and 
$\beta_O = 4 \exp[-(V+50)/18] = 83 \beta_B$ (ms$^{-1}$).
}
\end{center}
\end{figure*}

%Figure 10 
\begin{figure*}
\begin{center}
\includegraphics[width=0.7\textwidth]{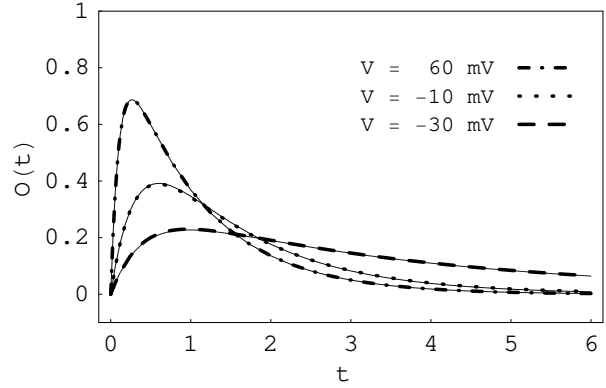}
\caption{
For a four state system where activation and inactivation are interdependent, during a  depolarizing clamp potential, the open state probability $O(t)$ (solid line)  may be approximated by  $m(t)h(t)$ 
(dashed, dotted or dot-dashed line) where $m(t)$  and $h(t)$ are solutions of rate equations
 for activation and inactivation, and  $\alpha_{1}(V) = \alpha_{2}(V) = 1$,  
$\gamma_{1}(V)= \gamma_{2}(V) = \exp(3)$, $\beta_{1}(V)= \beta_{2}(V) =  \exp(-2(V-10)/25)$ 
$\delta_{1}(V)= 4.2$, $\delta_{2}(V)= 0$, $\alpha_{O} = 0.1(V+25)/(1 - \exp[-(V+25)/10]) = \alpha_{B}/3$ and 
$\beta_O = 4 \exp[-(V+50)/18] = 83 \beta_B$ (ms$^{-1}$).
}
\end{center}
\end{figure*}

%Figure 11 
\begin{figure*}
\begin{center}
\includegraphics[width=0.7\textwidth]{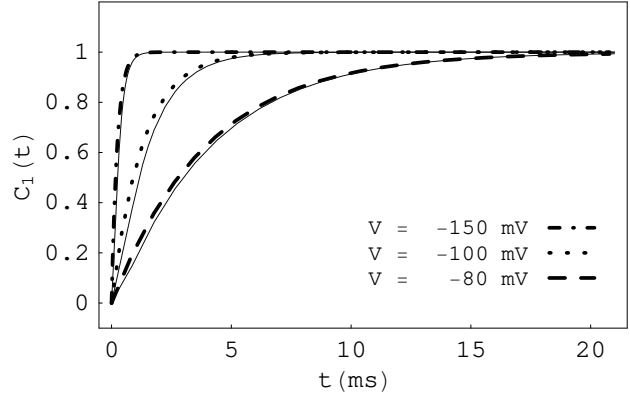}
\caption{
For a hyperpolarizing clamp potential of a four state system where activation and inactivation are interdependent, 
 the closed state variable  $C_1(t)$ (solid line) may be approximated by $[1 - m_s]h(t)$ (dashed, dot or dot-dashed)
 where $m_s = \alpha_{O}/(\alpha_{O} + \beta_O)$  and $h(t)$ is a solution of a rate equation for inactivation,
and rate functions are $\alpha_{1}(V) = \alpha_{2}(V) = 1$,  
$\gamma_{1}(V)= \gamma_{2}(V) = \exp(3)$, $\beta_{1}(V)= \beta_{2}(V) =  \exp(-2(V-10)/25)$ 
$\delta_{1}(V)= 4.2$, $\delta_{2}(V)= 0$, $\alpha_{O}  = 0.1(V+25)/(1 - \exp[-(V+25)/10]) = \alpha_{B}/3$ and 
$\beta_O = 4 \exp[-(V+50)/18] = 83 \beta_B$ (ms$^{-1}$).
}
\end{center}
\end{figure*}

%Figure 12 
\begin{figure*}
\begin{center}
\includegraphics[width=0.7\textwidth]{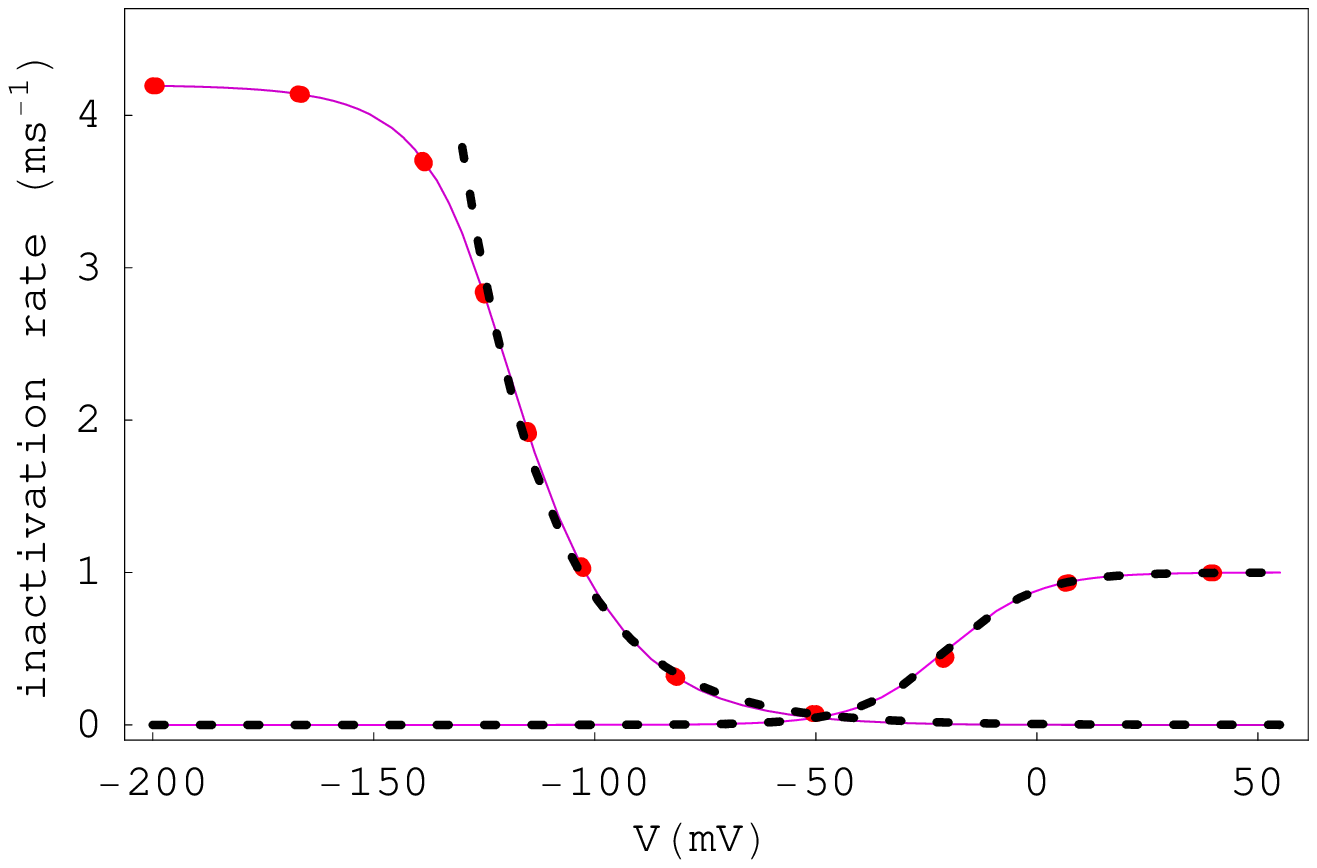}
\caption{
Voltage dependence of the HH Na+ channel inactivation rate functions $\beta_h = 1/(1+ \exp(-(20+V)/10))$ 
and $\alpha_h = 0.07 \exp(-(V+50)/20)$ (dashed line) may be approximated by analytical
 expressions (solid line) derived from a master equation  for a four state system 
where activation and  two stage inactivation  are interdependent, and by the voltage dependence
 of the lowest frequency of the system determined  numerically (dotted line) where the rate 
 functions are $\alpha_{1}(V) = \alpha_{2}(V) = 1$,  
$\gamma_{1}(V)= \gamma_{2}(V) = \exp(3)$, $\beta_{1}(V)= \beta_{2}(V) =  \exp(-2.5(V - 10)/25)$ 
$\delta_{1}(V)= 4.2$, $\delta_{2}(V)= 0$, $\alpha_{O} = \alpha_{B} = 5.4 \exp(-0.3V/25)$ and 
$\beta_O = 0.5 \exp[-V/18] = 98 \beta_B$ (ms$^{-1}$).
}
\end{center}
\end{figure*}

\end{document}